\documentclass[12pt]{article}
\usepackage{amsmath,amssymb}

\makeatletter
\@addtoreset{equation}{section}
\makeatother

\def\ben{\begin{equation}}
\def\een{\end{equation}}

  \let\n=\nu

\let\C=\Chi

\def\nn{\nonumber} \def\bd{\begin{document}} \def\ed{\end{document}}
\def\ds{\documentstyle} \let\fr=\frac \let\bl=\bigl \let\br=\bigr
\let\Br=\Bigr \let\Bl=\Bigl
\let\bm=\bibitem
\let\na=\nabla
\let\pa=\partial \let\ov=\overline
\newcommand{\be}{\begin{equation}}
\newcommand{\ee}{\end{equation}}
\def\ba{\begin{array}}
\def\ea{\end{array}}
\def\ft#1#2{{\textstyle{\frac{\scriptstyle #1}{\scriptstyle #2}}}}
\def\fft#1#2{\frac{#1}{#2}}

\def\del{\partial}
\def\vp{\varphi}
\def\sst#1{{\scriptscriptstyle #1}}
\def\oneone{\rlap 1\mkern4mu{\rm l}}
\def\td{\tilde}
\def\wtd{\widetilde}
\def\ie{\rm i.e.\ }
\def\dalemb#1#2{{\vbox{\hrule height .#2pt
        \hbox{\vrule width.#2pt height#1pt \kern#1pt
                \vrule width.#2pt}
        \hrule height.#2pt}}}
\def\square{\mathord{\dalemb{6.8}{7}\hbox{\hskip1pt}}}
\newcommand{\ho}[1]{$\, ^{#1}$}
\newcommand{\hoch}[1]{$\, ^{#1}$}
\newcommand{\bea}{\begin{eqnarray}}
\newcommand{\eea}{\end{eqnarray}}
\newcommand{\ra}{\rightarrow}
\newcommand{\lra}{\longrightarrow}
\newcommand{\Lra}{\Leftrightarrow}
\newcommand{\ap}{\alpha^\prime}
\newcommand{\bp}{\tilde \beta^\prime}
\newcommand{\tr}{{\rm tr} }
\newcommand{\Tr}{{\rm Tr} }
\def\0{{\sst{(0)}}}
\def\1{{\sst{(1)}}}
\def\2{{\sst{(2)}}}
\def\3{{\sst{(3)}}}
\def\4{{\sst{(4)}}}
\def\5{{\sst{(5)}}}
\def\6{{\sst{(6)}}}
\def\7{{\sst{(7)}}}
\def\8{{\sst{(8)}}}
\def\n{{\sst{(n)}}}
\def\cA{{{\cal A}}}
\def\cB{{{\cal B}}}
\def\cF{{{\cal F}}}
\def\tV{\widetilde V}
\def\tW{\widetilde W}
\def\tH{\widetilde H}
\def\tE{\widetilde E}
\def\tF{\widetilde F}
\def\tA{\widetilde A}
\def\im{{{\rm i}}}
\def\tY{{{\wtd Y}}}
\def\ep{{\epsilon}}
\def\vep{{\varepsilon}}
\def\R{\rlap{\rm I}\mkern3mu{\rm R}}
\def\bD{{{\bar D}}}

\def\R{\rlap{\rm I}\mkern3mu{\rm R}}
\def\bD{{{\bar D}}}
\def\R{{{\mathbb R}}}
\def\C{{{\mathbb C}}}
\def\H{{{\mathbb H}}}
\def\CP{{{\mathbb C}{\mathbb P}}}
\def\RP{{{\mathbb R}{\mathbb P}}}
\def\Z{{{\mathbb Z}}}
\def\bA{{{\mathbb A}}}
\def\bB{{{\mathbb B}}}
\def\bC{{{\mathbb C}}}
\def\bD{{{\mathbb D}}}
\def\bE{{{\mathbb E}}}
\def\bZ{{{\mathbb Z}}}
\def\Re{{{\frak{Re}}}}
\def\Im{{{\frak{Im}}}}
\def\cosec{{\,\hbox{cosec}\,}}
\def\Gm{{\Gamma_{\!\! -}}}
\def\Gp{{\Gamma_{\!\! +}}}
\def\stan{{standard }}
\def\nonstan{{supernumerary }}

\newcommand{\auth}{Chiang-Mei
Chen\hoch{\dagger} and Justin F. V\'azquez-Poritz\hoch{\ddagger}}

\begin{document}
\begin{flushright}

UK-04-06\ \ \ \ \
UCTP-106-04\\
March  2004\ \ \ \
\bf hep-th/0403109\\
\end{flushright}

\begin{center}
{\large {\bf Resolving the M2-brane}}

\vspace{10pt} \auth

\vspace{10pt} {\hoch{\dagger}\it Department of Physics,\\
National Central University, Chungli 320, Taiwan}

\vspace{10pt} {\hoch{\ddagger}\it Department of Physics and Astronomy,\\
University of Kentucky, Lexington, KY 40506}

\vspace{10pt} {\hoch{\ddagger}\it Department of Physics,\\
University of Cincinnati, Cincinnati OH 45221-0011}

\vspace{10pt}

\underline{ABSTRACT}
\end{center}

We construct deformed, $T^2$-wrapped, rotating M2-branes on a
resolved cone over $Q^{1,1,1}$ and $Q^{1,1,1}/\Z_2$, as well as on
a product of two Eguchi-Hanson instantons. All worldvolume
directions of these supersymmetric and regular solutions are
fibred over the transverse space. These constitute gravity duals
of $D=3$ $N=2$ gauge theories. In particular, the deformed
M2-brane on a resolved cone over $Q^{1,1,1}$ and the $S^1$-wrapped
M2-brane on a resolved cone over $Q^{1,1,1}/\Z_2$ provide explicit
realizations of holographic renormalization group flows in
M-theory for which both conformal and Lorentz symmetries are
broken in the IR region and restored in the UV limit. These
solutions can be dualized to supersymmetric type IIB pp-waves,
which are rendered non-singular either by additional flux or a
twisted time-like direction.



\pagebreak \setcounter{page}{1}


\newpage

\section{Introduction}

The AdS/CFT correspondence enables the study of strongly-coupled
gauge theories in terms of $p$-brane backgrounds
\cite{malda,gkp,wit}. In order to reduce the supersymmetry to a
minimum, one can replace the standard flat space that is
transverse to the $p$-brane by a space of special holonomy, which
is Ricci-flat and has fewer covariantly constant spinors. In order
for the solution to be non-singular, the transverse space must
have a non-collapsing $n$-cycle and thus support an $n$-form which
is square integrable at short distance. As a prelude to resolving
the M2-brane, we will briefly review the case of the well-studied
D3-brane of type IIB theory.

The simplest example of a six-dimensional Calabi-Yau manifold that
can be used for the transverse space is the conifold, defined as a
cone over $T^{1,1}=(S^3\times S^3)/S^1$. In the decoupling limit,
the resulting geometry is AdS$_5\times T^{1,1}$, which provides a
gravity dual to ${\cal N}=1$, $D=4$ superconformal Yang-Mills
theory. Wrapping 5-branes upon a supersymmetric 2-cycle in
$T^{1,1}$ gives rise to supersymmetric fractional D3-branes
\cite{kt,klebstra,ganpol,gubser}. A distance-dependent logarithmic
contribution to the D3-brane charge breaks the conformal symmetry.
However, this also induces a short-distance naked singularity and
therefore provides structural behavior only at large distance,
corresponding to the UV region of the dual gauge theory. The above
construction involves the complex 3-form $F_\3=F_\3^{{\rm
RR}}+i\,F_\3^{{\rm NS}}$, which is set proportional to a self-dual
3-form supported by the conifold. If the conifold has a
non-collapsing 3-cycle then the 3-form is square integrable at
short distance, resulting in the resolution of the above naked
singularity \cite{cglpns2d2}.

There are two smooth versions of the conifold, known as the
deformed conifold and the resolved conifold \cite{candel}. In the
former case, the singular apex is blown up to a smooth
three-sphere, yielding a non-collapsing 3-cycle. The
supersymmetric and regular fractional D3-brane on the deformed
conifold was constructed in \cite{klebstra}, and provides a
geometrical realization of chiral symmetry breaking and
confinement. On the other hand, for the resolved conifold, the
singular apex is blown up to a smooth two-sphere. Since it has a
collapsing 3-cycle, the fractional D3-brane over the resolved
conifold has a repulson-like naked singularity \cite{zaytse}. In
addition, it is not supersymmetric \cite{cglptrans,cglpsten}. The
same applies to the fractional D3-brane constructed with a
resolved cone over $T^{1,1}/\Z_2$ (also known as a regularized
conifold), for which the singular apex has been blown up to a
smooth $S^2\times S^2$ \cite{zaytse2}.

The non-collapsing 2-cycle of the resolved conifold over $T^{1,1}$
implies that there is a square integrable harmonic 2-form,
yielding a regular solution. Since there is no 2-form field
strength in type IIB theory, this contribution must take the form
of a fibration, as was originally observed for the D5-brane
\cite{lvres}. Thus, a D3-brane wrapped on $S^1$, which is fibred
over the resolved conifold, is regular and preserves a minimal
amount of supersymmetry \cite{wrapped}. In the case of the cone
over $T^{1,1}/\Z_2$, there is a non-collapsing 4-cycle. The square
integrable harmonic 4-form comes about by taking the Hodge dual of
a 2-form in the six-dimensional transverse space. Therefore, there
is a regular and supersymmetric $S^1$-wrapped D3-brane on the
resolved cone over $T^{1,1}/\Z_2$.  From the viewpoint of the dual
gauge theory, both conformal and Lorentz symmetries are broken in
the IR region and restored in the UV limit.

It has been conjectured that the fractional D3-brane and
$S^1$-wrapped D3-brane have a common origin in M-theory as a
modified supermembrane on a Spin(7) manifold, which gives rise to
the deformed and resolved conifolds in different Gromov-Hausdorff
limits \cite{wrapped}. The $G_2$ unification of deformed and
resolved conifolds supports this claim \cite{unif1,unif2,unif3}.

We will now turn to the case of the M2-brane, which has an
eight-dimensional transverse space. M2-branes with minimal
supersymmetries were first discussed in \cite{dlps} by replacing
the transverse $S^7$ with another Einstein space. Supersymmetric
and regular deformed M2-branes have been studied extensively, for
example, in \cite{cglpns2d2,cglptrans,cglpsten,beckers,deklm,
htr,becker,hk,cglphyper,cglpspin7,cgllp,gukov}. Singularity
resolution of the M2-brane solution due to gravitational
Chern-Simons corrections has been considered in \cite{brito}.

The main goal of this paper is the construction of supersymmetric
and regular $S^1$-wrapped M2-branes and rotating M2-branes
\cite{gauntlett}. One might wonder if this can be done indirectly
by dualizing the deformed M2-brane. Explicitly dualizing the
M2-brane to a type IIB pp-wave shows that this is not possible
from the type IIB context. On the other hand, one could imagine a
unification in higher dimensions, such as F-theory, analogous to
the M-theory unification of fractional and $S^1$-wrapped
D3-branes. Although the involvement of twelve dimensions is on
rather shaky ground, we will briefly consider this at least
formally, in the spirit of \cite{ftheory}. An immediate objection
would be that nine-dimensional Ricci-flat manifolds do not have an
irreducible special holonomy group under which eight-dimensional
manifolds of special holonomy can be unified.

Alternatively, there are eight-dimensional Ricci-flat K\"{a}hler
spaces which have both non-collapsing 4-cycles and non-collapsing
2-cycles, enabling the properties of a deformed M2-brane and
wrapped M2-brane to be combined into one solution. These spaces
include the cone over $Q^{1,1,1}$ and $Q^{1,1,1}/\Z_2$
\cite{castellani}, as well as the product space $M_4\times M_4$,
where each $M_4$ is either an Eguchi-Hanson \cite{eguchi} or
Taub-NUT instanton \cite{tn}. We find a self-dual harmonic
four-form and harmonic two-forms which are square integrable at
short distance. All of these components can be used to construct a
rather general supersymmetric and regular solution. This is a
deformed, $T^2$-wrapped and rotating M2-brane, since it is
modified by additional flux and all worldvolume directions,
including time, are fibred over the transverse space.

These solutions are dual to three-dimensional $N=2$ gauge
theories. In certain cases, the geometry asymptotically approaches
AdS$_4\times Q^{1,1,1}$, the CFT dual of which has been discussed
in \cite{hk,acharya,morrison,oh,dall,fre,ahn,fabbri,nunez}.
However, in the present solutions the singularity at the tip of
the conifold has been resolved, and the M2-brane solution is
rendered regular. These particular solutions exhibit holographic
renormalization group (RG) flows for which both conformal and
Lorentz symmetries are broken in general but are restored in the
UV limit. Other RG flows in M-theory which flow from
$E^{2,1}\times M_8$ at small distance to $AdS_4\times M_7$ at
large distance have been studied in \cite{nunez,hern}, which
correspond to $D=3$ super Yang Mills theories of various
supersymmetry.

For better clarity, we will first discuss deformed, wrapped and
rotating M2-branes separately, before combining them into the
general M2-brane solution. This paper is organized as follows. In
section 2, we discuss the deformed M2-brane on Ricci-flat
K\"{a}hler spaces. These M2-brane solutions are supersymmetric
and, due to additional flux, regular. In section 3, we discuss
$S^1$-wrapped or rotating M2-branes, for which the wrapped or
time-like coordinate of the worldvolume is fibred over the
transverse space. Now it is this fibration, rather than additional
flux, which results in the solution being regular. Since the
transverse spaces that we are considering support both an
appropriate harmonic 4-form and three harmonic 2-forms, we are
able to construct a rather general M2-brane in section 4. In
addition to being deformed by additional flux, this solution is
also $T^2$-wrapped and rotating. In section 5, we dualize the
general M2-brane to a type IIB supersymmetric pp-wave. The
geometry of the pp-wave is non-singular, due to both a variety of
fluxes as well as a twisted time-like direction. Similar pp-waves
are obtained from wrapped and rotating D3-branes and D5-branes.
The conclusions are given in section 6.

\section{Deformed M2-brane}

The M2-brane of eleven-dimensional supergravity is supported by
the 4-form field strength, with an eight-dimensional Ricci-flat
transverse space. Due to the equation of motion $d\ast F_\4=\ft12
F_\4\wedge F_\4$, one can construct a resolved M2-brane if the
transverse space has a (anti)-self-dual 4-cycle. This type of
modification to the M2-brane, which makes use of the interaction
in $d\ast F_\4=\ft12 F_\4 \wedge F_\4$, has been greatly studied,
for example in \cite{cglpns2d2,cglptrans,cglpsten,beckers,deklm,
htr,becker,hk,cglphyper,cglpspin7,cgllp}. The deformed M2-brane is
given by
\bea ds_{11}^2 &=& H^{-2/3}dx^{\mu}dx^{\nu}\eta_{\mu\nu}+H^{1/3}
ds_8^2\,,\nn\\
F_\4 &=& d^3x\wedge dH^{-1}+m\,G_\4\,, \label{deformed} \eea
where $G_\4$ is a harmonic self-dual 4-form in the Ricci-flat
transverse space $ds_8^2$. The equations of motion are satisfied,
provided that
\be \square H=-\fft{1}{48} m^2\,G_\4^2\,, \label{Heqn} \ee
where $\square$ is the Laplacian on $ds_8^2$. The introduction of
$G_\4$ to the M2-brane solution does not break additional
supersymmetries, provided that
\be G_{abcd}\Gamma^{bcd}\,\epsilon=0\,, \label{susy} \ee
where $\epsilon$ is a Killing spinor in the transverse space
\cite{becker}. The M2-branes discussed in this paper preserve
$\fft14$ of the original supersymmetry.

\subsection{On the resolved cone over $Q^{1,1,1}$}

For the transverse space, we will first consider an
eight-dimensional resolved cone over $Q^{1,1,1}$ (or
$Q^{1,1,1}/\Z_2$) \cite{castellani}, where the Einstein space
$Q^{1,1,1}$ can be expressed as the coset \cite{herzog}
\be Q^{1,1,1}=\fft{SU(2)^3}{U(1)^2}\,. \ee
The corresponding metric is of the form
\be ds_8^2 = d\rho^2 + c^2\,\sigma^2 + \sum_{i=1}^3\,a_i^2\,
(d\Omega_2^i)^2\,. \label{8metric} \ee
where
\be (d\Omega_2^i)^2 = d\theta_i^2+\sin^2 \theta_i\,
d\phi_i^2\,,\quad \sigma = d\psi+\sum_{i=1}^3\,\cos \theta_i\,
d\phi_i\,. \label{omega} \ee

The existence of Killing spinors implies the following first-order
equations:
\be \dot c=1-2\sum_i\,{\dot a}_i^2\,,\qquad c=2a_i\,{\dot a}_i\,,
\ee
where there is no implicit summation in the second expression.
These first-order equations imply that the space is Ricci-flat and
K\"ahler. The resolved eight-dimensional conifold is given by
\bea a_i^2 &=& \frac18(r^2+\ell_i^2)\,,\qquad c^2 =
\frac{r^2}{16h^2}\,,\qquad d\rho=h\,dr \nn \\
h^2 &=& \frac{3 \prod_i\,(r^2+\ell_i^2)}{3 r^6 + 4 \sum_i
\ell_i^2\,r^4 + 6\sum_{i<j}\ell_i^2 \ell_j^2\,r^2 + 12 \prod_i
\ell_i^2}\,. \label{8functions} \eea
This geometry with two non-vanishing $\ell_i$ was found in
\cite{cglpsten}, and with three non-vanishing $\ell_i$ was found
in \cite{cglpns2d2}. In each case, it was derived from the
first-order equations which result from the superpotential.

The range of the radial coordinate in is $0\le r< \infty$. In
order for the geometry to be regular at $r=0$, no more than one
$\ell_i$ can vanish. For two non-vanishing $\ell_i$, $\psi$ has a
period of $4\pi$ and the principal orbit is $Q^{1,1,1}$. The
geometry is $\R^4\times S^2\times S^2$ at small distance and the
cone over $Q^{1,1,1}$ asymptotically. Topologically, the manifold
is a $\C^2$ bundle over $\CP^1\times \CP^1$.

If all three $\ell_i$ are non-vanishing, then the period of $\psi$
is $2\pi$ and the principal orbit is $Q^{1,1,1}/Z_2$. The geometry
is $\R^2\times S^2\times S^2\times S^2$ at small distance and the
cone over $Q^{1,1,1}/\Z_2$ asymptotically. Topologically, the
manifold is a complex-line bundle over $\CP^1\times \CP^1\times
\CP^1$. If all three $\ell_i$ are equal, which will be a simple
case frequently discussed in this paper, then we can replace
$\CP^1\times \CP^1\times \CP^1$ by any other Einstein-K\"ahler
6-space. Note that the regular cone with three non-vanishing
$\ell_i$ does not reduce to the regular cone with two
non-vanishing $\ell_i$ in the limit that one of the $\ell_i$ goes
to zero, since the two spaces have different principal orbits.

The veilbein for the 8-space described by (\ref{8metric}) and
(\ref{8functions}) are given by
\be e^0=h\, dr\,, \qquad e^7 = c\,\sigma\,,\qquad e^{2j-1}=a_j\,
d\theta_j\,,\qquad e^{2j}=a_j\,\sin\theta_j \,d\phi_j\,, \ee
for $j=1,2,3$. We construct a harmonic self-dual 4-form on this
8-space with the ansatz
$$ G_\4 = u_1\,(e^0\wedge e^7\wedge e^1\wedge e^2+e^3\wedge
e^4\wedge e^5\wedge e^6)+u_2\,(e^0\wedge e^7\wedge e^3\wedge
e^4+e^1\wedge e^2\wedge e^5\wedge e^6) $$
\be +u_3\,(e^0\wedge
e^7\wedge e^5\wedge e^6+e^1\wedge e^2\wedge e^3\wedge e^4)\,. \ee
The closure of $G_\4$ yields
\bea u_1 &=& \fft{(\ell_2^2+\ell_3^2)(r^4+\ell_1^2\,\ell_2^2)
+2\ell_2^2
(\ell_1^2+\ell_3^2)r^2}{(r^2+\ell_1^2)(r^2+\ell_2^2)^2\,(r^2+\ell_3^2)^2}
\,,\nn\\ u_2 &=& -\fft{(\ell_1^2+\ell_3^2)(r^4+\ell_1^2\,\ell_2^2)
+2\ell_1^2(\ell_2^2+\ell_3^2)r^2}{(r^2+\ell_1^2)^2\,(r^2+\ell_2^2)
(r^2+\ell_3^2)^2}\,,\nn\\
u_3 &=& \fft{(\ell_1^2-\ell_2^2)(r^4-\ell_1^2\,\ell_2^2)}
{(r^2+\ell_1^2)^2\,(r^2+\ell_2^2)^2\,(r^2+\ell_3^2)}\,. \eea
For $\ell_1=0$, this reduces to the harmonic 4-form given in
\cite{cglpsten}. There is no charge associated with $G_\4$. Note
that the above three $u_i$'s satisfy the following linear
relation:
\be u_0+u_1+u_2=0\,, \label{u} \ee
which is a necessary, but not sufficient, condition for
supersymmetry. In particular, the supersymmetry condition
(\ref{susy}) implies (\ref{u}) but there are clearly four-forms
which satisfy the latter condition without giving rise to
supersymmetric solutions.

Since the general solution for $H$ is quite complicated, we will
focus on particular cases. For equal $\ell_i \equiv \ell$, we find
that
\bea H &=&
c_0+\fft{m^2}{\ell^4\,(r^2+\ell^2)}+\fft{m^2+c_1}{2\ell^6}\, {\rm
arctan} \Big( \fft{r^2+\ell^2}{\ell^2} \Big) \nn\\
&& +\fft{m^2-c_1}{4\ell^6}\, \log \Big( \fft{r^2}{r^2+2\ell^2}
\Big)\,, \eea
where $c_0$ and $c_1$ are arbitrary integration constants. In general,
there is a naked singularity as $r\rightarrow 0$. However, we can
choose $c_1=m^2$ such that the logarithmic term cancels, leaving a
non-singular solution given by
\be H=c_0+\fft{m^2}{\ell^4\,(r^2+\ell^2)}+\fft{m^2}{\ell^6}\, {\rm
arctan} \Big( \fft{r^2+\ell^2}{\ell^2} \Big)\,. \label{Hequal} \ee
We define $c_0$ so that $H$ asymptotically approaches unity at
large $r$. Also, this solution, as do all of the solutions
presented, has a well-defined ADM mass. The geometry interpolates
between $E^{4,1}\times S^2\times S^2\times S^2$ at small $r$ to
$E^{2,1}$ and a conifold at large $r$. Since the geometry does not
asymptotically approach AdS$_4$ times a compact space at large
distance, a dual gauge theory would not have conformal symmetry.

For the case of vanishing $\ell_1$, a regular solution is given by
\bea H &=&
c_0-\fft{3m^2\,(\ell_2^2+\ell_3^2+3r^2)}{2(2\ell_3^2-\ell_2^2)
(2\ell_2^2-\ell_3^2)(r^2+\ell_2^2)(r^2+\ell_3^2)}\nn\\
&& +\fft{27\sqrt{2}m^2}{4(2\ell_3^2-\ell_2^2)^{3/2}\,
(2\ell_2^2-\ell_3^2)^{3/2}}\, \arctan \Big[
\fft{\sqrt{2(2\ell_3^2-\ell_2^2)(2\ell_2^2-\ell_3^2)}}
{3r^2+2(\ell_2^2+\ell_3^2)}\Big]\,. \label{Hell1} \eea
This latter solution has already been found in \cite{cglpsten}.
For small $r$, the geometry becomes $E^{6,1}\times S^2\times S^2$.
For large $r$, and for an appropriate value of $c_0$, the geometry
asymptotically approaches AdS$_4\times Q^{1,1,1}$. This provides
an explicit realization of a holographic renormalization group
(RG) flow in M-theory for which both conformal and Lorentz
symmetries are broken in general but are restored in the UV limit
of the dual $D=3$ $N=2$ gauge theory. It is conjectured that,
since the M2-branes are coincident, the dual gauge theory is on
the Higgs branch \cite{cglpsten}. In general, the fact that there
is no charge associated with $G_\4$ implies that there are no
fractional branes. This indicates that the dual gauge theory is
perturbed by relevant operators \cite{hk}.

The M2-brane (\ref{deformed}) can be dimensionally reduced over
the transverse angular directions to give a $D=4$ domain-wall
solution, whose metric is given by
\be ds_4^2=H^{1/2}\,\prod_i a_i^2\,c\,(dx^{\mu} dx^{\nu}
\eta_{\mu\nu}+Hh^2\,dr^2)\,, \ee
where the various functions are given in (\ref{8functions}). For
vanishing $\ell_1$, $H$ is given by (\ref{Hell1}) and the
domain-wall metric asymptotically approaches AdS$_4$ at large $r$.
This implies that the solution is supported by a non-trivial
scalar potential in $D=4$ with a fixed point, which corresponds to
the holographic renormalization group flow of the dual field
theory \cite{cglpsten}.

\subsection{On $M_4\times M_4$}

We can use a product of two Ricci-flat K\"ahler 4-spaces, such as
the Eguchi-Hanson instanton \cite{eguchi} and Taub-NUT instanton
\cite{tn}, for the transverse space. We will consider two
Eguchi-Hanson instantons, for which the transverse metric is given
by
\be ds_8^2=ds_4^2+d{\tilde s}_4^2\,, \label{m4m4} \ee
where
\bea ds_4^2 &=& W^{-1}\,dr^2+\fft14
r^2\,W\,(d\psi+\cos\theta\,d\phi)^2
+\fft14 r^2\,(d\theta^2+\sin^2 \theta\,d\phi^2)\,,\nn\\
W &=& 1-\fft{a^4}{r^4}\,. \label{EHmetric} \eea
and likewise for $d{\tilde s}_4^2$. The radial coordinate $r$ lies
in the range $a\le r\le \infty$ and $\psi$ has the period $2\pi$.
The geometry is asymptotically locally Euclidean, and has level
surfaces of $S^3/\Z_2$ at constant $r$. The veilbein for the
8-space (\ref{m4m4}) are given by
\be e^0=W^{-1/2}dr\,,\quad e^1=\fft12 r\,d\theta\,,\quad
e^2=\fft12 r\,\sin\theta\,d\phi\,,\quad e^3=\fft12 r\,W^{1/2}
(d\psi+\cos\theta\,d\phi)\,, \ee
and likewise for ${\tilde e}^i$.

The Eguchi-Hanson metric $ds_4^2$ has a K\"ahler form given by
\be J_\2=(e^0\wedge e^3-e^1\wedge e^2)\,, \ee
and likewise for for $d{\tilde s}_4^2$ and ${\tilde J}_\2$. The
K\"ahler form for the 8-space (\ref{m4m4}) is given by
\be J=J_\2+{\tilde J}_\2\,. \ee
Also, $ds_4^2$ supports a self-dual harmonic 2-form given by
\be L_\2=\fft{2}{r^4}(e^0\wedge e^3+e^1\wedge e^2)\,, \label{L2}
\ee
and likewise for $d{\tilde s}_4^2$ and ${\tilde L}_\2$.

An anti-self-dual harmonic 4-form on the 8-space (\ref{m4m4}) can
be built out of the K\"ahler forms and harmonic 2-forms on the
individual Eguchi-Hanson spaces, given by
\be G_\4=L_\2\wedge {\tilde J}_\2+{\tilde L}_\2\wedge J_\2\,. \ee

The equation for $H$ (\ref{Heqn}) becomes
\be \Big( \fft{1}{r^3}\partial_r\, r^3\,W\,\partial_r+
\fft{1}{{\tilde r}^3}\partial_{{\tilde r}}\,{\tilde r}^3\, {\tilde
W}\,\partial_{{\tilde
r}}\Big)H=-8m^2\Big(\fft{1}{r^8}+\fft{1}{{\tilde r}^8}\Big)\,. \ee
For appropriate integration constants, a regular solution is given
by
\be H=1+\fft{m^2}{a^4\,r^2}+\fft{m^2}{{\tilde a}^4\,{\tilde
r}^2}\,. \label{m4m4H} \ee
The M2-brane geometry goes from $E^{6,1}\times S^2\times S^2$ at
short distance to $E^{10,1}$ at large distance. If one drops the
$1$ in the above $H$, then the solution is a domain-wall at large
distance.

For the eight-dimensional transverse space given by the metric
(\ref{m4m4}), we could also consider the product of two Taub-NUT
instantons, as well as the product of an Eguchi-Hanson and
Taub-NUT instanton. However, one has to take care that the spaces
are oriented such that both of the normalizable harmonic two-forms
are either both self-dual or anti-self-dual. This ensures that
cross-terms in functions of $r$ and ${\tilde r}$ cancel out in
$G_\4^2$, such that $H$ can easily be found in the simple form
$H=f(r)+g({\tilde r})$.

\section{Wrapped or rotating M2-brane}

If the transverse space has a 2-cycle $L_\2$, we can construct an
$S^1$-wrapped M2-brane with one of the world-volume coordinates
fibred over the transverse space.  Using the same technique
developed in \cite{lvres,wrapped,gauntlett}, we find that the
solution is given by
\bea ds_{11}^2 &=& H^{-2/3}\Big( -dt^2+dx_1^2+ (dx_2+A_\1)^2\Big)
+H^{1/3} ds_8^2\,,\nn\\
F_\4 &=& dt\wedge dx_1\wedge (dx_2+A_\1)\wedge dH^{-1}
+m\,dt\wedge dx_1\wedge L_\2\,, \label{wrappedm2} \eea
where $dA_\1=m\,L_\2$ and $L_\2$ is a harmonic 2-form in the
transverse space of the metric $ds_8^2$. The equations of motion
are satisfied if the function $H$ is given by
\be \square H=-\fft14 m^2\,L_\2^2\,, \label{H2} \ee
where $\square$ is the Laplacian on $ds_8^2$.

Alternatively, a rotating M2-brane solution is given by
\bea ds_{11}^2 &=& H^{-2/3}\Big( -(dt+A_\1)^2+dx_1^2+dx_2^2\Big)
+H^{1/3} ds_8^2\,,\nn\\
F_\4 &=& (dt+A_\1)\wedge dx_1\wedge dx_2\wedge dH^{-1}
+m\,dx_1\wedge dx_2\wedge L_\2\,, \label{rotatingm2} \eea
where again we require the function $H$ to satisfy (\ref{H2})
\cite{gauntlett}. Although the rotating M2-brane
(\ref{rotatingm2}) is related to the $S^1$-wrapped M2-brane
(\ref{wrappedm2}) by Wick rotations, this involves the analytical
continuation $m\rightarrow i\,m$ which results in a sign change in
(\ref{H2}). Although this can be counteracted by changing the
orientation of the conifold, reversing the orientation of the
transverse space can render the solution non-supersymmetric.

\subsection{On the resolved conifold over $Q^{1,1,1}$}

A general two-form on the 8-space given by (\ref{8metric}) and
(\ref{8functions}) can be expressed in terms of the veilbein as
\be L_{(2)} = u_0 \, e^0 \wedge e^7 + u_1\,e^1 \wedge e^2 + u_2\,
e^3 \wedge e^4 + u_3\,e^5 \wedge e^6\,, \ee
where $u_i$ are function of $r$ only. The K\"ahler form is given
by
\be J_\2=-e^0\wedge e^7+e^1\wedge e^2+e^3\wedge e^4+e^5\wedge
e^6\,. \ee
Its contribution to the function $H$ is badly behaved at large
distance and produces an unresolvable naked singularity. Also, it
breaks supersymmetry. Therefore, we will not consider the
contributions to $H$ from the K\"ahler form.

In addition to the K\"ahler form, there are three harmonic
2-forms. The first, which carries non-trivial flux, is given by
\bea u_0 &=& \frac{\alpha (r^2+\ell_3^2)^2 + 8 \beta
(r^2+\ell_3^2) + 3 \alpha
\beta}{(r^2+\ell_1^2)^2(r^2+\ell_2^2)^2}\,,\nn\\
u_1 &=& \frac{2 (r^2+\ell_3^2)^2 + 3 \alpha (r^2+\ell_3^2) + 6
\beta}{(r^2+\ell_1^2)^2(r^2+\ell_2^2)}\,,
\nn\\
u_2 &=& \frac{2 (r^2+\ell_3^2)^2 + 3 \alpha (r^2+\ell_3^2) + 6
\beta}{(r^2+\ell_1^2)(r^2+\ell_2^2)^2}\,,\nn\\
u_3 &=& - \frac{ 4 (r^2+\ell_3^2) + 3
\alpha}{(r^2+\ell_1^2)(r^2+\ell_2^2)}\,, \label{form1} \eea
where $\alpha=\ell_1^2+\ell_2^2-2\ell_3^2$ and $\beta = \ell_1^2
\ell_2^2 + \ell_3^4 - \ell_1^2 \ell_3^2 - \ell_2^2\ell_3^2$.

Since it is difficult to solve for the most general $H$ from
(\ref{H2}), we will consider a few specific cases for which the
solution is regular. For $\ell_3=0$,
\be H=c_0 +
\frac{3m^2(\ell_1^2-3\ell_2^2)^2}{8(\ell_1^2-\ell_2^2)(\ell_1^2-2\ell_2^2)\,
(r^2+\ell_1^2)} +
\frac{3m^2(\ell_2^2-3\ell_1^2)^2}{8(\ell_2^2-\ell_1^2)(\ell_2^2-2\ell_1^2)\,
(r^2+\ell_2^2)} \ee $$ -\frac{27 m^2 (\ell_1^2-\ell_2^2)^2}{8
(\ell_1^2-2\ell_2^2) (\ell_2^2-2\ell_1^2)
\sqrt{2(\ell_1^2-2\ell_2^2) (\ell_2^2-2\ell_1^2)}} \arctan\Big[
\frac{3r^2+2\ell_1^2+2\ell_2^2}
{\sqrt{2(\ell_1^2-2\ell_2^2)(\ell_2^2-2\ell_1^2)}} \Big]\,, $$
where the integration constant $c_0$ is chosen such that $H$
asymptotically approaches unity for large $r$.

For the case of all equal $\ell_i \equiv \ell$, we find that a
regular $H$ is given by
\be H=c_0-\fft{3m^2}{2\ell^2}\,{\rm arctan}\Big(
\fft{r^2+\ell^2}{\ell^2} \Big)\,. \label{H1} \ee

The second harmonic 2-form also carries non-trivial flux and is
given by
\bea u_0 &=& \frac{2(\ell_2^2-\ell_3^2)(r^2+\ell_3^2)^2 + 8 \beta
(r^2+\ell_3^2) + [ 3\alpha +\tilde\alpha ]
\beta}{(r^2+\ell_1^2)^2(r^2+\ell_2^2)^2}\,,\nn\\
u_1 &=& \frac{2(\ell_2^2-\ell_3^2)(r^2+\ell_3^2) +4
\beta}{(r^2+\ell_1^2)^2(r^2+\ell_2^2)}\,,\nn\\
u_2 &=& \frac{4(r^2+\ell_3^2)^2 + [5\alpha -\tilde\alpha ]
(r^2+\ell_3^2) +8\beta}{(r^2+\ell_1^2)(r^2+\ell_2^2)^2}\,,\nn\\
u_3 &=& - \frac{ 4(r^2+\ell_3^2) + [3\alpha +\tilde\alpha
]}{(r^2+\ell_1^2)(r^2+\ell_2^2)}\,, \label{form2} \eea
where $\tilde\alpha=\ell_1^2-\ell_2^2$. A regular solution for
$\ell_3=0$ is given by
\be H=c_0+
\fft{3m^2\ell_2^4}{2(\ell_1^2-\ell_2^2)(\ell_1^2-2\ell_2^2)\,
(r^2+\ell_1^2)} +
\fft{3m^2(\ell_2^2-2\ell_1^2)}{2(\ell_2^2-\ell_1^2)\,
(r^2+\ell_2^2)} \ee $$ -\fft{3 m^2 (\ell_2^2-2\ell_1^2)}{2
(\ell_1^2-2\ell_2^2) \sqrt{2(\ell_1^2-2\ell_2^2)
(\ell_2^2-2\ell_1^2)}} \arctan\left(
\frac{3r^2+2\ell_1^2+2\ell_2^2}
{\sqrt{2(\ell_1^2-2\ell_2^2)(\ell_2^2-2\ell_1^2)}} \right). $$

For the case of all equal $\ell_i \equiv \ell$, we find that a
regular $H$ has the same form as (\ref{H1}) with $m\rightarrow
2m/\sqrt{3}$.

The last harmonic 2-form has vanishing flux, namely
\be \int_{r\rightarrow \infty} L_\2=0\,. \ee
This 2-form is given by
\bea u_0 &=& \frac{3r^4 + 2\sum_i\ell_i^2\,r^2 +\sum_{i\ne j
}\ell_i^2\ell_j^2}
{(r^2+\ell_1^2)^2(r^2+\ell_2^2)^2(r^2+\ell_3^2)^2}\,,\nn\\
u_1
&=& \frac1{(r^2+\ell_1^2)^2(r^2+\ell_2^2)(r^2+\ell_3^2)}\,,\nn\\
u_2 &=&
\frac1{(r^2+\ell_1^2)(r^2+\ell_2^2)^2(r^2+\ell_3^2)}\,,\nn\\ u_3
&=& \frac1{(r^2+\ell_1^2)(r^2+\ell_2^2)(r^2+\ell_3^2)^2}\,.
\label{form3} \eea
The resulting solution is only regular if all $\ell_i$ are
non-vanishing.

For the case of all equal $\ell_i \equiv \ell$, we find that a
regular $H$ is given by
\be H=1+\fft{m^2}{8\ell^8\,(r^2+\ell^2)^3}\,. \label{solution} \ee
The previous solutions of this section have geometries which
interpolate between a product space of $E^{3,1}$ and a $U(1)$
bundle over $S^2\times S^2\times S^2$ at short distance to
$E^{2,1}$ and a conifold at large distance. Therefore, if there is
a dual gauge theory, then it does not have a phase for which
conformal symmetry is restored. On the other hand, the solution
given in (\ref{solution}) is particularly interesting since, if we
drop the $1$, the metric interpolates from a product space of
$E^{1,1}$ and a $U(1)$ bundle over $\R^2\times S^2\times S^2\times
S^2$ at short distance to $AdS_4\times Q^{1,1,1}/\Z_2$ at large
distance. As with the deformed M2-brane on a cone over
$Q^{1,1,1}$, this solution exhibits a holographic renormalization
group (RG) flow for which both conformal and Lorentz symmetries
are broken in general but are restored in the UV limit of the dual
$D=3$ $N=2$ gauge theory. Since there is no charge associated with
the 4-form term which arises from the 2-form (\ref{form3}), this
indicates that the dual gauge theory is perturbed by relevant
operators \cite{hk}.

Note that, for all three harmonic 2-forms, the four $u_i$'s
satisfy the following linear relation:
\be u_0 - u_1 - u_2 - u_3 = 0\,. \ee
This is a necessary, but not sufficient, condition for
supersymmetry. For example, the K\"{a}hler form satisfies the
above relation without satisfying the supersymmetry equation
(\ref{susy}).

\subsection{On $M_4\times M_4$}

The 2-form $L_\2$, given by (\ref{L2}), is not only harmonic on
the Eguchi-Hanson metric (\ref{EHmetric}) but on the
eight-dimensional metric (\ref{m4m4}) as well. Likewise for the
2-form ${\tilde L}_\2$. Therefore, we can construct the following
harmonic 2-form on (\ref{m4m4}):
\be G_\2=L_\2+{\tilde L}_\2\,. \ee
For appropriate integration constants, this yields a regular $H$
given by (\ref{m4m4H}) with $m\rightarrow m/\sqrt{2}$. As with the
deformed M2-brane, we can also find a regular wrapped or rotating
M2-brane on a product of two Taub-NUT instantons or a product of
an Eguchi-Hanson and Taub-NUT instantons.

\section{General M2-brane}

A deformed, $T^2$-wrapped, rotating M2-brane on a resolved
conifold over $Q^{1,1,1}/\Z_2$ can be constructed which contains
all of the elements of the previous two sections, by using both
the harmonic 4-form and the harmonic 2-forms. An analogous
M2-brane can also be constructed on $M_4\times M_4$, where each
$M_4$ is either an Eguchi-Hanson or Taub-NUT space. However, the
latter case only gives us two harmonic two-forms at our disposal,
which means we must decide between either a deformed,
$T^2$-wrapped (non-rotating) M2-brane or a deformed,
$S^1$-wrapped, rotating M2-brane. All of these general M2-branes
are supersymmetric and regular.

This is the only known cases of a deformed and wrapped/rotating
$p$-brane solution which is supersymmetric and regular. For
example, although both fractional D3-branes and wrapped D3-branes
have been constructed, in order to be supersymmetric and regular,
different transverse spaces are required. The fractional D3-brane
requires a deformed conifold which supports a harmonic 3-form
while the wrapped (or rotating) D3-brane requires a resolved
conifold which supports harmonic 2-forms. However, a
supersymmetric and regular $T^2$-wrapped and $S^1$-wrapped
rotating D3-brane can be constructed from two independent harmonic
2-forms supported by the resolved conifold, which we will discuss.

In the case of the D5-brane, four-dimensional transverse spaces
such as the Eguchi-Hanson and Taub-NUT instantons only support a
single harmonic 2-form for which the solution is supersymmetric
and regular solution. Multi-instanton spaces are expected to
support multiple independent 2-forms which would presumably give
rise to deformed, wrapped, rotating D5-branes which are
supersymmetric and regular. These have yet to be constructed.
Also, a Schwarzschild instanton does support two independent
harmonic 2-forms but while the resulting D5-brane solutions are
regular, they are not supersymmetric.

A deformed, $T^2$-wrapped, rotating M2-brane is given by
\bea ds_{11}^2 &=& H^{-2/3}\,d{\tilde x}^{\mu}d{\tilde
x}^{\nu}\eta_{\mu \nu}
+H^{1/3} ds_8^2\,,\nn\\
F_\4 &=& d^3{\tilde x}\wedge dH^{-1} +m_4\,G_\4+ \sum_{i\ne j\ne
k} m_k\,d{\tilde x}^i\wedge d{\tilde x}^j\wedge L_\2^k \,,
\label{generalm2} \eea
where $d{\tilde x}^i\equiv dx^i+A_\1^i$ and $i,j,k=0,1,2$. As an
example, we will consider the transverse space to be the resolved
conifold over $Q^{1,1,1}/\Z_2$. $G_\4$ is the harmonic 4-form
found in section 2.1. $dA_\1^i=m_i\,L_\2^i$ ($i$ not summed),
where $L_\2^i$ are the three harmonic 2-forms found in section
2.1. The equations of motion are satisfied, provided that
\be \square H=-\fft{1}{48}m_4^2\,G_\4^2-\fft14 \sum_{i=0}^2 m_i^2
(L_\2^i)^2\,. \ee

We consider the case of all equal $\ell_i\equiv \ell$. A regular
$H$ is given by
\be H=c_0+\fft{m_4^2}{\ell^4\,(r^2+\ell^2)}
+\fft{m_0^2}{8\ell^8\,(r^2+\ell^2)^3} + \Big( \fft{m_4^2}{\ell^6}
- \fft{3m_1^2}{2\ell^2} - \fft{2m_2^2}{\ell^2} \Big) \, \arctan
\Big( \fft{r^2+\ell^2}{\ell^2}\Big)\,, \ee
which is just a superposition of individual $H$ for each harmonic
forms. The roles of $m_0$, $m_1$ and $m_2$ are, of course,
interchangeable. Likewise, a general M2-brane can be constructed
on the product of Eguchi-Hanson or Taub-NUT instantons. For the
case of two Eguchi-Hanson instantons, the resulting regular $H$
has the same form as in (\ref{m4m4H}).

\section{Resolved pp-waves}

\subsection{From M2-branes}

The general M2-brane given by (\ref{generalm2}) can be
dimensionally reduced on $x_2$ to give a deformed, wrapped,
rotating NS-NS string in type IIA theory, given by
\bea ds_{10}^2 &=& H^{-3/4}\,( -d{\tilde x}_0^2+d{\tilde x}_1^2)
 +H^{1/4} ds_8^2\,,\nn\\
F_\4 &=& m_4\,G_\4+m_2\,d{\tilde x}_0\wedge d{\tilde x}_1\wedge
L_\2^2\,,\nn\\
F_\3 &=& d{\tilde x}_0\wedge d{\tilde x}_1\wedge dH^{-1} +m_0\,
d{\tilde x}_1\wedge L_\2^0+m_1\,d{\tilde x}_0\wedge L_\2^1\,,\nn\\
F_\2 &=& m_2\, L_\2^2\,,\qquad e^{2\phi}=H\,. \label{generalns1}
\eea
Hopf T-duality on the $x_1$ coordinate has the effect of
untwisting this direction \cite{hopf}. At the same time, the first
two terms in the above 3-form flux are dualized to contributions
to the metric. The result is a supersymmetric and regular type IIB
pp-wave given by
\bea ds_{10}^2 &=& -H^{-1}\,d{\tilde x}_0^2+H\,\Big(
dx+B_\1 \Big)^2+ds_8^2\,,\nn\\
F_\5 &=& m_4\,(dx+B_\1)\wedge (G_\4+\ast_8 G_\4)\,,\nn\\
F_\3^{{\rm RR}} &=& m_2\,(dx+d{\tilde x}_0+B_\1)\wedge L_\2^2\,,\nn\\
F_\3^{{\rm NS}} &=& m_1\,(dx+d{\tilde x}_0+B_\1)\wedge L_\2^1\,,
\label{generalm2pp} \eea
where $(H^{-1}-1)\,d{\tilde x}_0$. If instead we had reduced the
M2-brane over $x_1$ and Hopf T-dualized over $x_2$, then we would
have ended up with the S-dual of the pp-wave solution
(\ref{generalm2pp}), with $F_\3^{{\rm RR}}\leftrightarrow
F_\3^{{\rm NS}}$.

This general solution encompasses various limits in which one or
more of the fields vanish, which are related to deformed, wrapped
or rotating M2-branes. The connection between deformed M2-branes
and supersymmetric type IIB pp-waves has been discussed in
\cite{m2pp}. As long as all of the fields $L_\2^i$ and $G_\4$ do
not vanish simultaneously, the solution (\ref{generalm2pp}) is
completely regular. In particular, vanishing $L_\2^1$, $L_\2^2$
and $G_\4$ yields a Ricci-flat pp-wave, whose twisted time-like
direction ensures a non-singular geometry.

Finally, for vanishing $L_\2^2$ and $G_\4$, (\ref{generalns1}) is
also a solution of type IIB theory, which can be S-dualized to a
wrapped, rotating D1-brane. This solution can then be Hopf
T-dualized to a (smeared) rotating D0-brane. All of these
solutions are supersymmetric and regular.

Note that this is more fortunate than the situation for the
deformed D1-brane on the conifold, which has a singular geometry.
This is because the transverse space does not have a
non-collapsing 5-cycle at short distance, and therefore cannot
admit an appropriate 5-form \cite{cglpns2d2,herzog}.

\subsection{From D3-branes}

An $S^1$-wrapped rotating D3-brane is given by \cite{wrapped}
\bea ds_{10}^2 &=& H^{-1/2}\,\Big(
-(dt+A_\1^0)^2+dx_1^2+dx_2^2+(dx_3+A_\1^1)^2
\Big) +H^{1/2}\,ds_6^2\,,\nn\\
F_\5 &=& (dt+A_\1^0)\wedge dx_1\wedge dx_2\wedge
(dx_3+A_\1^1)\wedge
dH^{-1}\\
&& +m_0\,\ast_6 L_\2^0\wedge (dt+A_\1^0)+m_1\,\ast_6 L_\2^1 \wedge
(dx_3+A_\1^1)+{\rm dual\,\, terms}\,,\nn \eea
where $dA_\1^i=m_i\,L_\2^i$ for $i=0,1$. $L_\2^i$ are two
independent harmonic 2-forms on $ds_6^2$, and $\ast_6$ is the
Hodge dual with respect to $ds_6^2$. The equations of motion are
satisfied, provided that
\be \square H=-\fft12 \sum_{i=0}^1\,m_i^2 (L_\2^i)^2\,,
\label{eqnH3} \ee
where $\square$ is the Laplacian on $ds_6^2$. A regular and
supersymmetric solution exists for the case in which $ds_6^2$ is
the metric for the resolved conifold on $T^{1,1}/Z_2$. For the
case in which the D3-brane is not rotating, the metric
interpolates between a product of $E^{2,1}$ and a $U(1)$ bundle
over $\R^2\times S^2\times S^2$ at short distance to AdS$_5\times
T^{1,1}/\Z_2$ at large distance. This provides a regular gravity
dual to a certain four-dimensional gauge theory whose Lorentz and
conformal symmetries are broken in the IR region and restored in
the UV limit \cite{wrapped}.

Multiple T-dualities yields a type IIB pp-wave, which is given by
\bea ds_{10}^2 &=& -H^{-1}\,(dt+A_\1^0)^2+H\,\Big(
dx+(H^{-1}-1)(dt+A_\1^0) \Big)^2+ds_6^2+dz_1^2+dz_2^2\,,\nn\\
F_\5 &=& m_1\,\ast_6 L_\2^1\wedge dx+m_1\,dz_1\wedge dz_2\wedge
L_\2^1+{\rm dual\,terms}\,. \eea
Notice that, for vanishing $L_\2^1$, the solution is Ricci-flat,
for which the twisted time-like direction ensures that the
geometry is regular.

Alternatively, we could consider a $T^2$-wrapped D3-brane, which
dualizes to a type IIB pp-wave given by
\bea ds_{10}^2 &=& -2dt\,dx+H\,dx^2+ds_6^2+dz_1^2+dz_2^2\,,\nn\\
F_\5 &=& m_1\,\ast_6 L_\2^1\wedge dx+m_1\,dz_1\wedge dz_2\wedge
L_\2^1+{\rm dual\,terms}\,,\nn\\
F_\3^{{\rm RR}} &=& m_0\,(dx+dt)\wedge L_\2^0\,. \eea
Dualizing the fibred directions in reverse order results in the
S-dual of the above pp-wave solution, with an NS-NS 3-form.

On the other hand, if $ds_6^2$ is a deformed conifold, then there
is a regular and supersymmetric fractional D3-brane supported by a
self-dual harmonic 3-form $L_\3$. This solution can be dualized to
a type IIB pp-wave given by
\bea ds_{10}^2 &=& -2dt\,dx+H\,dx^2+ds_6^2+dz_1^2+dz_2^2\,,\nn\\
F_\5 &=& m\,dx\wedge (dz_1+dz_2)\wedge L_\3+{\rm dual\,terms}\,.
\eea

\subsection{From D5-branes}

The $S^1$-wrapped D5-brane is given by \cite{lvres}
\bea ds_{10}^2 &=& H^{-1/4}\,\Big( -dt^2+dx_1^2+\cdots +dx_4^2
+(dx_5+A_\1)^2\Big) +H^{3/4}ds_4^2\,,\nn\\
F_\3^{{\rm RR}} &=& \ast_4 dH-m\,L_\2\wedge (dx_5+A_\1)\,, \qquad
\phi=-\fft12\,\log H\,, \eea
where $dA_\1=m\,L_\2$. $L_\2$ is a harmonic 2-form on $ds_4^2$ and
$\ast_4$ is the Hodge dual with respect to $ds_4^2$. The equations
of motion are satisfied, provided that
\be \square H=-\fft14 m^2\,L_\2^2\,, \label{eqnH5} \ee
where $\square$ is the Laplacian on $ds_4^2$. There exist regular
and supersymmetric solutions for the case in which $ds_4^2$ is the
metric for the Eguchi-Hanson or Taub-NUT instanton. In the case of
the Eguchi-Hanson (Taub-NUT) instanton, the geometry of the
D5-brane goes from a product of $E^{4,1}$ and a $U(1)$ bundle over
$\R^2\times S^2$ ($\R^4$) at short distance to a domain-wall at
large distance.

These solutions can be dualized to type IIB pp-waves, given by
\bea ds_{10}^2 &=& -2dt\,dx+H\,dx^2+ds_6^2+dz_1^2+dz_2^2\,,\nn\\
F_\5 &=& m\,dx\wedge (dz_1\wedge dz_2+dz_3\wedge dz_4)\wedge
L_\2+{\rm dual\,terms}\,. \eea

We can also consider a rotating D5-brane, given by
\bea ds_{10}^2 &=& H^{-1/4}\,\Big( -(dt+A_\1)^2+
dx_1^2+\cdots +dx_5^2 \Big)+H^{3/4}ds_4^2\,,\nn\\
F_\3^{{\rm RR}} &=& \ast_4 dH-m\,L_\2\wedge (dt+A_\1)\,, \qquad
\phi=-\fft12\,\log H\,, \eea
where again we require the function $H$ to satisfy (\ref{eqnH5}).
This solution can be dualized to a pp-wave, given by
\be ds_{10}^2=-H^{-1}\,(dt+A_\1)^2+H\,\Big( dx+(H^{-1}-1)(dt+A_\1)
\Big)^2+ds_4^2+dz_1^2+\cdots +dz_4^2\,. \ee

\section{Conclusions}

We have constructed various supersymmetric and regular M2-brane
solutions. For the transverse space, we have used
eight-dimensional Ricci-flat K\"{a}hler spaces which have both
non-collapsing 4-cycles and non-collapsing 2-cycles. Such spaces
include a resolved cone over $Q^{1,1,1}$ and $Q^{1,1,1}/\Z_2$, as
well as a product of two Eguchi-Hanson or Taub-NUT instantons. The
non-collapsing 4-cycle enables the construction of a square
integrable, self-dual, harmonic four-form on the transverse space,
which results in a deformed M2-brane. A non-collapsing 2-cycle
means that the transverse space supports a square integrable,
harmonic two-form, which takes the form of a fibration in an
$S^1$-wrapped or rotating M2-brane. Combining all of these
elements into a single solution yields a deformed, $T^2$-wrapped,
rotating M2-brane. All of the worldvolume directions, including
time, are fibred over the transverse space.

Dualizing this M2-brane yields a supersymmetric and regular
pp-wave in type IIB theory. Interestingly enough, in the limit in
which all fluxes vanish, the geometry of the Ricci-flat pp-wave is
still non-singular, due to the twisted time-like direction.
Additional pp-waves of this class are obtained from wrapped and
rotating D3-branes and D5-branes.

The M2-brane solutions are dual to three-dimensional $N=2$ gauge
theories. The geometry of the deformed M2-brane on a resolved
conifold over $Q^{1,1,1}$ smoothly goes from $E^{6,1}\times
S^2\times S^2$ at short distance to AdS$_4\times Q^{1,1,1}$ at
large distance. Also, the $S^1$-wrapped M2-brane on a resolved
conifold over $Q^{1,1,1}/\Z_2$ has a geometry that goes from a
product space of $E^{1,1}$ and a $U(1)$ bundle over $\R^2\times
S^2\times S^2\times S^2$ at short distance to $AdS_4\times
Q^{1,1,1}/\Z_2$ at large distance. These particular solutions
exhibit holographic renormalization group flows for which both
conformal and Lorentz symmetries are broken in general but are
restored in the UV limit. There are also cases, such as the
deformed M2-brane on a resolved cone over $T^{1,1}/\Z_2$ and a
deformed or wrapped M2-brane on $M_4\times M_4$, for which the
geometry approaches a domain-wall at large distance and conformal
symmetry is not restored in the UV limit of the dual gauge theory.

Unlike typical brane solutions which require an M-theoretic brane
source term, resolved branes are complete purely within
supergravity. However, not all BPS branes can be resolved at the
level of supergravity. One benefit of $S^1$-wrapped $p$-branes is
that they can still be non-singular even when a deformed or
fractional $p$-brane on the same transverse space has a
singularity. For example, a fractional D-string on a resolved cone
over $Q^{1,1,1}/\Z_2$ does not have regular short-distance
behavior. This is because the transverse space does not have a
non-collapsing 5-cycle from which to construct a 5-form which is
square integrable at short distance. It was suggested that either
non-perturbative string effects were required to resolve the
singularity \cite{cglpns2d2} or else it is cloaked by a horizon at
sufficiently high temperature \cite{herzog}. On the other hand, we
were able to dualize our $S^1$-wrapped M2-brane to an
$S^1$-wrapped D1-brane in type IIB theory, which is non-singular.
This is because the transverse space does have a non-collapsing
2-cycle, which yields an appropriate fibration.

As an earlier example, we consider the resolution of the type II
5-brane. A regular heterotic or type II 5-brane wrapped around the
$S^2$ of a resolved conifold was constructed in \cite{mn}.
Alternatively, the heterotic 5-brane on an Eguchi-Hanson or
Taub-NUT instanton can be resolved by making use of multiple
matter Yang-Mills fields which are absent in the type II theories
\cite{cglptrans}. For this reason, the analogous resolution of the
type II 5-brane was unknown before the construction of the
$S^1$-wrapped 5-brane \cite{lvres}, which makes use of a fibration
over the transverse space.

The list of known supersymmetric and regular $S^1$-wrapped
$p$-branes on spaces of special holonomy includes all $0\le p\le
5$. However, since there are no irreducible nine and
five-dimensional manifolds of special holonomy, the D0 and
D4-brane must have transverse product spaces. For example, we
could use $M_n\times S^1$, where $M_n$ is an irreducible
$n$-dimensional manifold of special holonomy, and $n=8$ for the
D0-brane and $n=4$ for the D4-brane \cite{cglptrans}. Likewise, a
regular $S^1$-wrapped D2-brane can be easily constructed on
$M_6\times S^1$. However, a regular $S^1$-wrapped D2-brane on a
$G_2$ holonomy space, if there is one with a non-vanishing
2-cycle, has yet to be constructed.

\section*{ACKNOWLEDGMENT}

It is a pleasure to thank Adel Awad and Hong L\"{u} for helpful
correspondence, as well as Matthew Strassler for a useful
conversation. The work of C.M.C. is supported in part by the
National Science Council of the R.O.C. under grant number
NSC92-2119-M-008-024. The work of J.F.V.P. is supported in part by
DOE grant DE-FG01-00ER45832.

\end{document}